%% file: main.tex
\documentclass{article}

\usepackage{amsmath}
\usepackage{verbatim}
\usepackage{graphicx}
\usepackage{amssymb}
\usepackage[authoryear]{natbib}
\usepackage{authblk}

\usepackage{setspace}

\DeclareMathOperator*{\argmin}{argmin}

\newcommand{\R}{\mathbb{R}}
\newcommand{\U}{\mathcal{U}}
\newcommand{\X}{\mathcal{X}}
\newcommand{\B}{\mathcal{B}}
\newcommand{\Rg}[1]{\text{Rgrt}_\U \, (#1)}
\newcommand{\PRg}[1]{\text{PRgrt}_\U \, (#1)}

\graphicspath{{/}}

\begin{document}

\input{Intro}
\input{Survey}
\input{Model}

\input{numeric}

\bibliography{refs}
\bibliographystyle{plainnat}

\input{appendix}

\end{document}

%% file: Intro.tex
\title{Robust Portfolio Optimisation with Specified Competitors}

\author[1,2]{Gon\c{c}alo Sim\~{o}es}
\author[3]{Mark McDonald}
\author[ ]{Stacy Williams}
\author[3]{Daniel Fenn}
\author[1,2]{Raphael Hauser}
\date{}

\affil[1]{Mathematical Institute, University of Oxford, OX2 6GG, UK}
\affil[2]{Oxford-Man Institute of Quantitative Finance, University of Oxford, OX2 6ED, UK}
\affil[3]{FX Quantitative Research, HSBC Bank, London E14 5HQ, UK}

{\let\newpage\relax\maketitle}

\begin{abstract}
We extend Relative Robust Portfolio Optimisation models to allow portfolios to optimise their distance to a set of benchmarks. Portfolio managers are also given the option of computing regret in a way which is more in line with market practices than other approaches suggested in the literature. In addition, they are given the choice of simply adding an extra constraint to their optimisation problem instead of outright changing the objective function, as is commonly suggested in the literature. We illustrate the benefits of this approach by applying it to equity portfolios in a variety of regions.
\end{abstract}

Modern Portfolio Theory has been an area of active research in mathematical finance since \cite{markowitz}, but it has not been fully adopted by practitioners yet. One of its known shortfalls is the assumption that future returns' moments are known with certainty, which has led to substantial under-performance in practice, see for example \cite{demiguel-garlappi-uppal,michaud}. Recently however there have been developments made to try to tackle this issue. Robust Optimisation is one of the techniques used to tackle this problem after having been successfully applied to other fields, see \cite{bertsimas-brown-caramanis}.

The fact that model parameters (such as future returns' moments) are unknown is the core motivation behind Robust Optimisation. Instead of specifying with certainty what the model parameters are, Robust Optimisation models only require a set of possible parameter values, the so-called uncertainty set. Another popular technique, Stochastic Optimisation, requires not only an uncertainty set but also an associated probability distribution.

By far the most studied version of Robust Portfolio Optimisation is the worst-case scenario approach, whose aim is to find the portfolio with the best performance under the worst possible values in the uncertainty set. A large number of papers have been published with varied extensions being proposed, from different shapes of uncertainty sets, different market model assumptions to inclusion of transaction costs. For a survey see \cite{kim-kim-fabozzi}.

While this approach may be sensible in some situations, we believe this not well suited for most practitioners. Even though it is important to worry about extreme scenarios, typical everyday scenarios are no less important and should not be ignored. Moreover, some professionals such as investment managers are frequently evaluated against the competition and not on absolute terms. For those reasons we believe that Relative Robust Optimisation, introduced in \cite{kouvelis-yu} and developed in \cite{hauser-tutuncu}, is more suited to most portfolio managers. In this approach a portfolio's worth is assessed not only on its performance but also on the competition's performance. Relative Robust Optimisation has not been as widely studied as its worst-case scenario counterpart because its appeal lies mainly in financial applications and not in engineering, the birthplace of Robust Optimisation.

Competition is clearly an ambiguous term which needs to be made precise. In \cite{hauser-tutuncu} the authors define competition as an ``omniscient adversary'' that has the same constraints as we do, but knows the ``correct'' model parameters and hence solves a non-robust portfolio optimisation problem. We introduce the idea that competition may be specified independently of our initial problem, i.e., competition may not be bound by the same constraints we are, nor is it using the same techniques. We make it clear that competition may be described according to user's needs, just like the uncertainty set in classical Robust Optimisation problem is. If competition happens to be bound by exactly the same constraints as we are and is solving an optimisation problem, then we recover the setup in \cite{hauser-tutuncu}.

Most academic literature on Robust Portfolio Optimisation propose that the objective function must be replaced, which suggests investors must start anew if they wish to use such a tool (and they can never use more than one simultaneously). However investors have their own objectives and are (with reason) wary of changing them, which makes the leap from literature to practice much harder to achieve. With this in mind we propose instead adding an extra constraint, not a new objective function. This allows the investor to keep his framework intact and simply make a minor change, trading an extra constraint for extra robustness.

For simplicity, but also for practical purposes, we will use a finite set as our competition. This can be extended to a finite collection of more general convex sets, using standard tools from Convex Analysis. However we would caution practitioners to be wary of overcomplicating their models without having a deep understanding of the potential shortcomings. When in doubt, simpler is better.

In line with this philosophy we will look at models that deal with volatility, not expected returns. The idea of investing without taking into account expected returns is not new. Since \cite{haugen-baker} proposed a minimum volatility portfolio a lot of research has been published on it and its performance has been widely studied, see for example \cite{clarke-silva-thorley,lee}. In fact, some asset managers already offer this solution to their clients in one form or another (see \cite{scherer}) and it was recently reported in the media that such funds had net inflows of $\$12.5$ billion in the first half of 2016 (see  \cite{WSJ}). Extensions, so called risk-based investing, have been developed, including maximum diversification (\cite{choueifaty-coignard}), equal risk contribution (\cite{maillard-roncalli-teiletche}), among others (\cite{jurczenko-michel-teiletche}). 

It is however straightforward to extend our framework to models that require expected returns or some other feature to be provided. In fact this approach appears particularly promising given how sensitive traditional Portfolio Optimisation is to errors in the means, see \cite{chopra-ziembra}. Covariance matrices in comparison are much more stable over time, although they do shift between different regimes, as shown in \cite{RORO}. We will assume a finite number of regimes, each characterised by a covariance matrix. We then demonstrate that our problem is extendible to the case where the uncertainty set, the set of all undetermined model parameters, is within a class of general convex sets.

When introducing a new model an important thing to look for is whether the result is fundamentally different than simpler models already present in the literature. Using equities data from a range of regions we employ the methodology in \cite{changepoint} to investigate possible regime switching in the covariance matrix. We run our model against some commonly used approaches and we conclude that the resulting portfolio is not only distinct but a sensible alternative to existing methods.

We start with a short review of relevant existing models: first the minimum volatility problem, followed by absolute robust portfolio optimisation and finally relative robust portfolio optimisation. We then move on to introduce our own model, by proposing a different measure of \textit{regret} than the one found in  \cite{hauser-tutuncu}. Initially we consider it an objective function, as is common in the literature, but subsequently we propose using it as a constraint instead, making it more widely applicable. We then pursue the suggestion in \cite{kouvelis-yu} and introduce \textit{proportional regret} in the context of portfolio optimisation, a slightly different measure that is more in line with practitioners common practice. At last we run the numerical experiments described above and we conclude with some final remarks.

It is worth mentioning that all the proposed changes and additions do not alter the complexity of the underlying optimisation problem. Second order cone programs or semidefinite programs will still be within their respective classes after Relative Robust Optimisation has been implemented and so can still be solved in polynomial time using a standard solver such as MOSEK or CPLEX).

%% file: Survey.tex
\section*{Classical Models}

We shall start by describing the most common models found in the literature before putting forward a different approach to deal with uncertainty in the covariance matrix.

Let $\X$ be the set of admissible portfolios, which we assume for simplicity to be the set of all long-only portfolios,
\begin{align} \label{simple}
\X = \{x \in \R^N : 1^T x = 1, x \geq 0\}.
\end{align}
These are not at all required and extra constraints may be added due to user's preference and/or regulatory restrictions.

We start with the simple problem of finding the minimum volatility portfolio when we know the ``true'' covariance matrix $Q$, introduced in \cite{haugen-baker}.  This problem can be written as
\begin{align} \label{minvol}
\min_x \quad & \sqrt{x^T Q \, x} \\
\text{s.t.} \quad & x \in \X. \notag
\end{align}

This model takes $Q$ as known, which is not a valid assumption in financial applications. While there are many techniques available to provide good estimates, any approach of practical relevance is going to have to deal with the fact that market conditions change, as described in \cite{RORO}, and so assuming a covariance matrix as certain is bound to fail. A natural step forward is to assume that the future covariance matrix is an unknown element of a set of \textit{scenarios} $\U = \{Q_1, \dots, Q_n\}$.

The challenge is how to make a decision when given such an uncertainty set. One possible approach is to assign some probability to each of the scenarios in $\U$ and then minimise volatility weighted by each scenario, as follows
\begin{align*}
\min_x \quad & \sum_{i=1}^n p_i \sqrt{x^T Q_i \, x} \\
\text{s.t.} \quad & x \in \X.
\end{align*}

This is not a very attractive option, as it is already non-trivial to find a sensible set $\U$, let alone somehow pick an ``appropriate'' probability distribution.

Another common approach is \textit{Absolute Robust Portfolio Optimisation}, which has been extensively studied in the literature, see for example \cite{kim-kim-fabozzi}. The aim is to minimise volatility on the worst-case scenario, i.e., on the scenario where volatility is highest for the chosen portfolio. This can be represented as
\begin{align} \label{absolute}
\min_x \; \max_{Q \in \U} \quad & \sqrt{x^T Q \, x} \\
\text{s.t.} \quad & x \in \X. \notag
\end{align}

The problem is that this model focuses \textbf{only} on the worst-case scenario. Therefore this approach completely ignores all the other scenarios, leading to a portfolio that while being fairly stable during stress periods may be very unbalanced if these do not come to pass.

A more recent approach is the one found in \cite{hauser-tutuncu}, \textit{Relative Robust Portfolio Optimisation}. The goal here is not to minimise volatility, but instead to minimise the amount of extra volatility we are forced to accept for not knowing the covariance matrix with certainty. To put it in mathematical terms, for each $Q \in \U$ define $y_Q$ as
\begin{align*}
y_Q := \argmin_{y \in \X} \; \sqrt{y^T Q \, y}, \quad \forall Q \in \U,
\end{align*}
i.e., if we knew $Q_k$ to be the ``true'' scenario then we should hold $y_{Q_k}$ (by solving (\ref{simple})). For any other portfolio $x$ we might choose, our loss in performance would be
\begin{align*}
l(x,Q_k) := & \sqrt{x^T Q_k \, x} - \sqrt{y_{Q_k}^T Q_k \, y_{Q_k}} \\
& \sqrt{x^T Q_k \, x} - \min_{y \in \X} \; \sqrt{y^T Q_k \, y}.
\end{align*}

But we do not know which $Q \in \U$ is going to be realised and so we cannot know how much our loss in performance will be. We can however strive to make it as low as possible, by minimising the worst outcome across all possible values of $Q \in \U$, that is, by solving
\begin{align} \label{relative}
\min_x \; \max_{Q \in \U} \quad & \left( \sqrt{x^T Q \, x} - \min_{y \in \X} \; \sqrt{y^T Q \, y} \right) \\
\text{s.t.} \quad & x \in \X. \notag
\end{align}

%% file: Model.tex
\section*{Regret Minimisation}

While this method has potential, we wish to approach it differently. Comparing us against a ``more knowledgeable version of ourselves'' might be a very interesting theoretical problem, but in practice one gets compared against other players in the market, not fictional beings.  Hence we cannot be satisfied by the formulation in (\ref{relative}).

We would like instead to be compared against other investors, as this would be a more realistic portrait of the real world. Unfortunately we not know what other agents in the market are investing in. Their constraints are not the same as our own -- and even if they were, it is implausible all of them are solving (\ref{simple}). We cannot therefore hope to model precisely the behaviour of all investors.

As an alternative, we will compare ourselves against benchmarks, which are naturally unambiguous. Benchmarks have two properties that make them suitable for this purpose. On the one hand, an asset manager is not only compared against better performing competitors, but also against relevant benchmarks. This means comparison against benchmarks is a reasonable representation of reality.

\begin{figure}[htb]
\includegraphics[width=\textwidth]{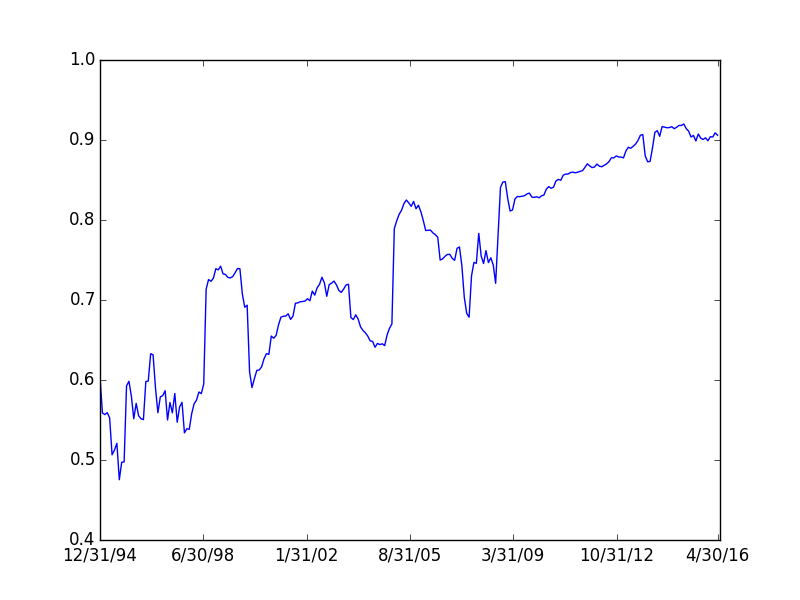}
\caption{Correlation between the HFRI Equity Hedge Index and the S{\&}P 500} \label{fig::hedge}
\end{figure}

On the other hand, competitors will themselves be victims of pressure not to fall too short of a particular benchmark. This will in turn pull them towards a similar basket, resulting in a high correlation between the two. 
See for example the historical correlation between the HFRI Equity Hedge Total Index and the S{\&}P 500 in Figure \ref{fig::hedge}. This chart provides support to the claim that benchmarks are a good proxy for agents' returns. We can see in this particular example that hedge funds' returns have been increasingly correlated with the  S{\&}P 500, with recent values surpassing the $90\%$ threshold.

This is the case despite the fact that most hedge funds do not have explicit benchmarks -- their mandate is to deliver positive returns and they are commonly considered a diversified investment.

\subsection*{Modelling Regret}

Let $\B = \{b_1, \dots, b_k\} \subseteq \R^N$ be the set of benchmarks we wish to consider. Following the rationale behind (\ref{relative}), we wish to consider our loss in performance to the least volatile benchmark,
\begin{align*}
l_\B(x,Q) := \sqrt{x^T Q \, x} - \min_{b \in \B} \; \sqrt{b^T Q \, b}.
\end{align*}
Here we assumed that all the underlyings of all benchmarks are included in the set of $N$ assets considered for investment, and so we can compute their volatility by $\sqrt{b^T Q \, b}$ -- in truth, we only need to specify the volatility of the least volatile benchmark under each scenario.

In a similar fashion to \cite{hauser-tutuncu} we define \textit{regret} to be the maximum of this loss over $\U$, i.e.,
\begin{align} \label{regret}
\Rg{x} := \max_{Q \in \U} l_\B(x,Q) = \max_{Q \in \U} \left( \sqrt{x^T Q \, x} - \min_{b \in \B} \; \sqrt{b^T Q \, b} \right) .
\end{align}
Regret can be understood as our distance to the ``winner'' in the least desirable scenario.

Benchmarks choice, just like scenarios, is beyond the scope of this paper. We can however provide a few useful tips that apply to both. First off, they should be picked independently of the portfolio optimisation problem, be it through economic/political considerations, data analysis, or any other technique. Furthermore, they should be as realistic and as few as possible. Too many scenarios and benchmarks will lead to overly defensive portfolios, since an ``anything can happen'' attitude is a sure way to investment paralysis.

\subsection*{Regret Minimisation as an Objective}

If we wish to find the portfolio that has least regret then we have to solve the following \textit{Relative Robust Portfolio Optimisation with Benchmarks} problem.
\begin{align} \label{minregret}
\min_x \quad & \Rg{x} \\
\text{s.t.} \quad & x \in \X. \notag
\end{align}
Unlike the minimum volatility problem (\ref{minvol}), minimum regret \textit{may} be negative. It should be clear to the reader that this will be the case if and only if there exists a portfolio that has a lower volatility than any benchmark regardless of which scenario is realised.

As mentioned before, we do not need the volatility of every benchmark for every scenario -- we simply require the lowest benchmark volatility for each scenario. Hence we define the following
\begin{align*} 
\sigma_i = \min_{b \in \B} \; \sqrt{b^T Q_i \, b} \quad i=1,\dots,n.
\end{align*}

Problem (\ref{minregret}) can then be untangled into 
\begin{align*}
\min \; & \gamma \\
\text{s.t.} \quad & \gamma \geq \sqrt{x^T Q_i \, x} - \sigma_i, \quad \forall i=1,\dots,n , \\
& x \in \X.
\end{align*}

The above optimisation problem can be converted into a second order conic program, which can in turn be efficiently solved using standard solvers, by introducing auxiliary variables $t_i$. The resulting optimisation problem is
\begin{align}
\min \; & \gamma \label{SOCP} \\
\text{s.t.} \; & \gamma - t_i + \sigma_i \geq 0 , \quad \forall i=1,\dots,n ,\notag  \\
& \begin{bmatrix}
U_i^T x \\
t_i
\end{bmatrix} \in L_{N+1}, \quad \forall i=1,\dots,n , \notag \\
& x \in \X, \notag
\end{align}
where $Q_i = U_i \, U_i^T$ is the Cholesky decomposition and $L_{N+1}$ is the Lorentz cone defined by
\begin{align*}
L_{N+1} = \left\{ \begin{bmatrix}
x \\
t
\end{bmatrix} \in \R^{N+1} :
\| x \|_2 \leq t \right\}.
\end{align*}


\subsection*{Regret as a Constraint}

So far we have only seen models that assume volatility as the objective function. Another common approach is to use volatility instead as a constraint. In fact \cite{markowitz} first introduced Modern Portfolio Theory as
\begin{align*}
\max_x \quad & \mu^T \, x \\
\text{s.t.} \quad & \sqrt{x^T Q \, x} \leq \kappa,
\end{align*}
where $\mu$ is the expected future returns and $\kappa$ the level of volatility we are prepared to accept in our portfolio. $\kappa$ can thus be as low as the solution of (\ref{minvol}).

Similarly, we can look at using regret not as an objective in itself, but rather as a constraint to be imposed in our portfolio allocation problem. Say for example we wish to maximise expected return as in the above problem.  Then we could solve a similar problem,
\begin{align} \label{muregret}
\max_x \quad & \mu^T \, x \\
\text{s.t.} \quad & \Rg{x} \leq \theta, \notag
\end{align}
where $\theta$ is the regret we are prepared to accept and can be no lower than the solution to (\ref{minregret}).

This approach is attractive on the grounds that investors prefer to target more meaningful objectives, such as maximising expected returns or minimising transaction costs. In addition it gives the portfolio manager control of how much risk he is prepared to take in order to increase (or decrease) his objective function. It is also a more flexible option, since adding a constraint is comparatively easier than rewriting from scratch an optimisation problem with a new objective function.

\subsection*{Proportional Regret}

As an alternative to regret we could define loss of performance not as the difference in volatilities, but instead the proportion of benchmark volatility we surpass, i.e.,
\begin{align*}
l'_\B(x,Q) := \frac{\sqrt{x^T Q \, x} - \min_{b \in \B} \sqrt{b^T Q \, b}}{\min_{b \in \B} \sqrt{b^T Q \, b}},
\end{align*}
We believe thinking in proportions is more natural for investors and so this option will be of more practical interest to them.

By using $l'_\B$ we can define the notion of \textit{proportional regret}
\begin{align*}
\PRg{x} := \max_{Q \in \U}  \, \max_{b \in \B} \,  \frac{\sqrt{x^T Q \, x} -  \sqrt{b^T Q \, b}}{\sqrt{b^T Q \, b}}.
\end{align*}
As before, we only require the volatility of the best benchmark under each scenario to be modelled, not for all of them. Likewise, proportional regret can only be negative if $x$ outperforms all benchmarks over all scenarios.

It should be made clear however that this is not just a cosmetic change. While the best benchmark in each scenario remains the same, the trade-off between scenarios does change. To illustrate this let us look at a quick example.

Imagine we have two assets we can invest in, but because of external constraints we can only invest in one of them. We are however aware that we will be compared against a single benchmark -- an equally weighted basket of the two assets. Our research team concluded that over the next year there are two possible scenarios, summarised in these two covariance matrices:

\begin{align*}
Q_1 = \left.\begin{bmatrix}
0.2 & 0.085 \\
0.085 & 0.23
\end{bmatrix}\right. \quad \quad Q_2 = \left.\begin{bmatrix}
0.18 & 0.18 \\
0.18 & 0.26
\end{bmatrix}\right.
\end{align*}

The first asset has 20pp\footnote{To avoid confusion we will refer to volatility units (usually $\%$) as pp.} volatility in Scenario 1 and 23pp volatility in Scenario 2, while the second asset has 18pp and 26pp volatility in Scenario 1 and 2, respectively. In addition we have a single benchmark that has 15pp volatility in Scenario 1 and 20pp volatility in Scenario 2.

One can check that $x$ performs worst in Scenario 1, while $y$ performs worst in Scenario 2. $x$ trails $b$ by 5pp, or $33.3\%$, while $y$ trails $b$ by 6pp, or $30\%$. It is then clear that if we cared about regret we would invest in $x$, while if we cared about proportional regret we would invest in $y$. Hence these two approaches to measuring distance to benchmarks yield different decisions.

Proportional regret can be used as an objective simply by substituting $\text{Rgrt}_\U$ by $\text{PRgrt}_\U$ in (\ref{minregret}). It can also be used as a constraint, in which case we make the same substitution in (\ref{muregret}) instead.

%% file: numeric.tex
\section*{Empirical Results}

In this section we will investigate if the proposed method yields sensible results and, equally important, if the results are fundamentally different from other commonly used approaches.

To do this we will investigate its use in building a long-only portfolio of  equities across different regions. Since our goal is mostly illustrative we choose to regard as investable assets industry group indices instead of individual companies. For that purpose we follow the recommendations of \cite{GICS} and use the Global Industry Classification Standard (GICS), a proprietary classification standard jointly produced by Standard $\&$ Poor's and Morgan Stanley Capital International. We will therefore consider each of the 24 industry groups in GICS as investable assets.

Benchmark choice is a delicate process that will vary substantially from one investor to the next. Given this, we will take, for simplicity's sake, advantage of the hierarchical structure of GICS and use as benchmarks its 10 sectors along with the region wide total return index. We will use data from 2001 onwards, with prices quoted in local currency, and for each specific region will only consider industry groups and sectors that had a non-interrupted presence.

Another choice that needs to be made is what are the scenarios we wish to consider. There exist an almost countless number of ways to generate scenarios that take into account all kinds of data and/or expertise. We made the decision to solely use historical data up to 2012 in order to generate our scenarios, which we will then use to compute portfolios under different portfolio optimisation techniques. Data from 2013 to April 2016 will then be used to analyse these portfolios' performance. The scenarios will be covariance matrices of subperiods of our training horizon. To determine the changepoints we implement the algorithm proposed in \cite{changepoint}. This is an attractive option since it makes no distribution assumption and so fits well in the spirit of these experiments. While we could in theory have ``forced'' some dates to be changepoints (Lehman Brothers collapse or 9/11, just to name a couple possibilities), we wish to have as little direct influence as possible and so will stick with the dates generated by the algorithm.

\subsection*{Minimum Regret in the EMU}

\begin{figure}[!htb]
\includegraphics[width=\textwidth]{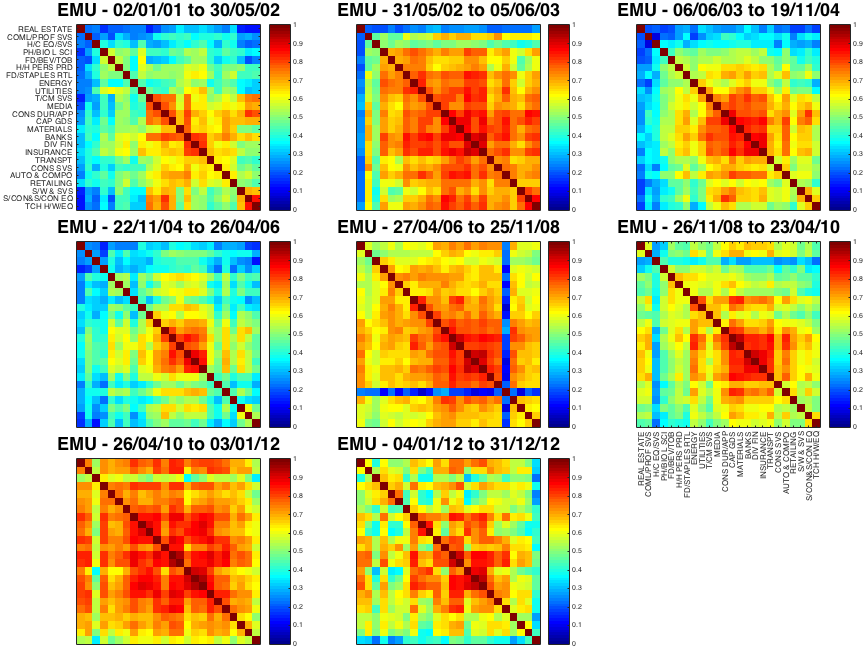}
\caption{Heatmaps of correlation matrix for different scenarios in the EMU. Ordering is preserved throughout all matrices.} \label{fig::EMU_Heat}
\end{figure}

We will start by focusing on the European Economic and Monetary Union equity market. After running the algorithm proposed in \cite{changepoint}, we arrive at the following changepoints: 31/5/02, 06/06/03, 22/11/04, 27/04/06, 26/11/08, 26/04/10 and 04/01/12. From these we construct 8 sub-periods, which we then summarise into 8 covariance matrices to be our scenarios. We can visualise them through the correlation heatmaps in Figure \ref{fig::EMU_Heat}.

\begin{figure}[!htb]
\includegraphics[width=\textwidth]{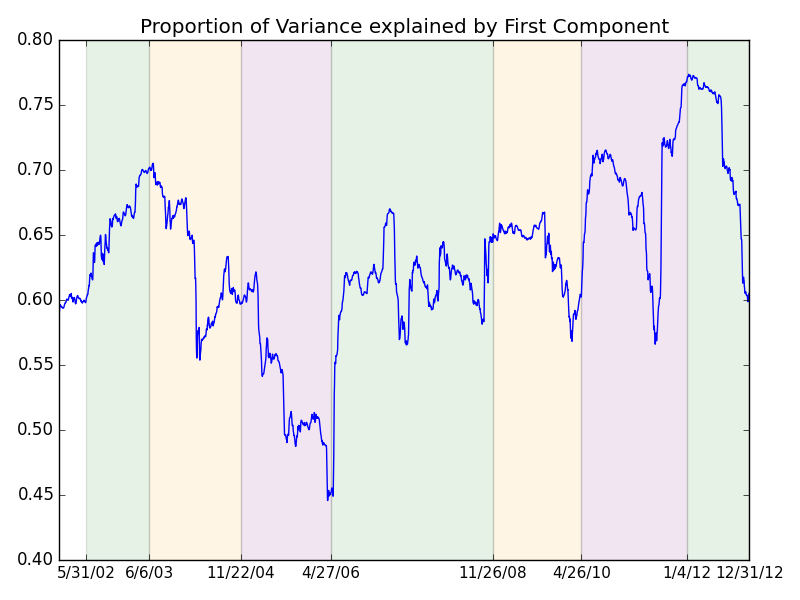}
\caption{Proportion of Variance explained by First Principal Component in EMU over the different scenarios.} \label{fig::EMU_RORO}
\end{figure}

We check whether these scenarios are sensible by following the methodology proposed in \cite{RORO}. In Figure \ref{fig::EMU_RORO} we can see the behaviour of the first principal component of a one-year rolling window of returns over different scenarios.

This graph does seem to suggest that the first component strength changed quite significantly across different scenarios, which is in line with our assumption that these constitute different regimes.

We are now in a position to run our portfolio (RRPO), as in (\ref{minregret}), and compare it to other methods. Two methods that are reasonable to compare our model against are the absolute robust portfolio optimisation (ARPO), as in (\ref{absolute}), using the same scenarios as RRPO, and a simple minimum volatility portfolio optimisation (MVPO), as in (\ref{simple}), using the whole training period to compute the covariance matrix estimate. These portfolios have their weights detailed in Table \ref{table::EMU_comp}.

\begin{table}[!htb]
\begin{center}
\begin{tabular}{| l | r | r | r | r |}
\hline
& RRPO & ARPO & MVPO \\ \hline
Food, Beverage \& Tobacco & 35.84\% & 32.96\% & 36.77\% \\ \hline
Health Care Equipment \& Services & 33.28\% & 50.23\% & 44.17\% \\ \hline
Food \& Staples Retailing & 10.15\% & 0\% & 0\% \\ \hline
Media & 10.13\% & 0\% & 0\% \\ \hline
Telecommunication Services & 5.74\% & 0\% & 2.16\% \\ \hline
Real Estate & 4.86\% & 4.09\% & 12.26\% \\ \hline
Transportation & 0\% & 9.65\% & 4.64\% \\ \hline
Automobiles \& Components & 0\% & 3.07\% & 0\% \\ \hline
\end{tabular}
\end{center} \caption{Weights of Relative Robust Portfolio Optimisation (RRPO), Absolute Robust Portfolio Optimisation (ARPO) and Minimum Volatility Portfolio Optimisation (MVPO) for EMU.} \label{table::EMU_comp}
\end{table}

Upon examining this table, one can immediately draw a few conclusions. First, all three portfolios concentrate on a small subset of all investable options (GICS has 24 industry groups, whereas Table \ref{table::EMU_comp} makes mention of only 8 of them). This is hardly surprising, as all of them are aiming at the same objective: having low volatility. It is then only natural that they tend to concentrate on low volatility assets. Second, it looks like RRPO is slightly more diversified than ARPO and MVPO, with a maximum weight of $35.84\%$ over 6 assets against a maximum of $50.23\%$ and $44.17\%$ over 5 of the other methods. This can be quantified by the Gino coefficient\footnote{The Gino coefficient is a measure of dispersion, with 0 representing equal dispersion (a 1/N allocation) and 1 representing highly unbalanced allocations.} -- RRPO has a Gino coefficient of 0.849, while ARPO and MVPO have a Gino coefficient of 0.894 and 0.889 respectively. Finally, it is evident that RRPO is yielding a structurally different portfolio to both ARPO and MVPO -- the proportion of capital allocated differently is $29.67\%$ and $23.86\%$ respectively, both non-trivial quantities.

\begin{figure}[!htb]
\includegraphics[width=\textwidth]{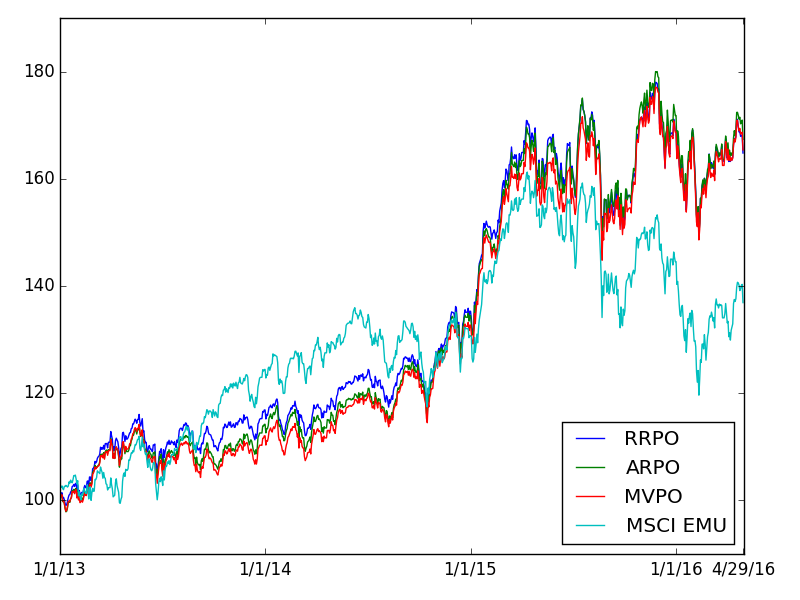}
\caption{Total return of RRPO, ARPO, MVPO and MSCI EMU} \label{fig::EMU_perf}
\end{figure}

Let us now have a look at how these portfolios perform out of sample, over the period starting from 2013 until April 2016. In Figure \ref{fig::EMU_perf} we can see the total return progress over time while in Table \ref{table::EMU_perf} we compile the yearly realised return and volatility of each portfolio.

\begin{table}[!htb]
\begin{center}
\begin{tabular}{| l | r | r | r | r |}
\hline
& RRPO & ARPO & MVPO & MSCI EMU \\ \hline
Realised Return 2013 & 15.04\% & 12.80\% & 10.97\% & 21.09\% \\ \hline
Realised Return 2014 & 14.13\% & 15.79\% & 16.34\% & 4.78\% \\ \hline
Realised Return 2015 & 21.45\% & 23.66\% & 23.19\% & 9.70\% \\ \hline
Realised Return 2016 & -7.25\% & -7.71\% & -5.82\% & -15.93\% \\ \hline
Realised Return 2013-16 & 14.49\% & 14.93\% & 14.59\% & 9.11\% \\ \hline
Realised Volatility 2013 & 11.95pp & 12.33pp & 12.23pp & 14.29pp \\ \hline
Realised Volatility 2014 & 12.36pp & 13.10pp & 12.72pp & 15.41pp \\ \hline
Realised Volatility 2015 & 20.94pp & 21.55pp & 21.45pp & 21.21pp \\ \hline
Realised Volatility 2016 & 21.53pp & 22.59pp & 21.98pp & 24.14pp \\ \hline
Realised Volatility 2013-16 & 16.33pp & 16.95pp & 16.73pp & 18.05pp \\ \hline
\end{tabular}
\end{center} \caption{Annualised realised return and volatility of RRPO, ARPO, MVPO and MSCI EMU over different periods.} \label{table::EMU_perf}
\end{table}

It so turns out that for this dataset all three methods behave remarkably similarly, even if one takes into account the fact that roughly $75\%$ of capital is allocated equally. As we will see later on, this behaviour is purely coincidental -- the same methods across different regions have discernible differences in realised return.

\subsection*{Maximum Expected Return in the US}

We now wish to test another of the proposed methods, a proportional regret constraint of $10\%$ in a maximum expected return portfolio optimisation (PRCPO), as in (\ref{muregret}) but with $\text{PRgrt}_\U$ instead of $\text{Rgrt}_\U$. This will give us an example of a situation where the objective is already set, and we solely add a constraint to force a more stable behaviour. We will compare this against a Sharpe ratio absolute robust portfolio optimisation (SARPO) (where we aim to have the best worst-case Sharpe ratio) and against a maximum Sharpe ratio portfolio optimisation (MSPO). Moreover we will see how these methods compare when there are other restrictions present. We will not only keep the long-only constraint, but we will also add a $20\%$ cap to force diversification.

We will use US equity market data so we can look in detail at another dataset. Our investable assets will still be the 24 GICS industry groups and our benchmarks the 10 GICS sector groups plus the MSCI US index. The training and testing period also remain unchanged

\begin{figure}[!htb]
\includegraphics[width=\textwidth]{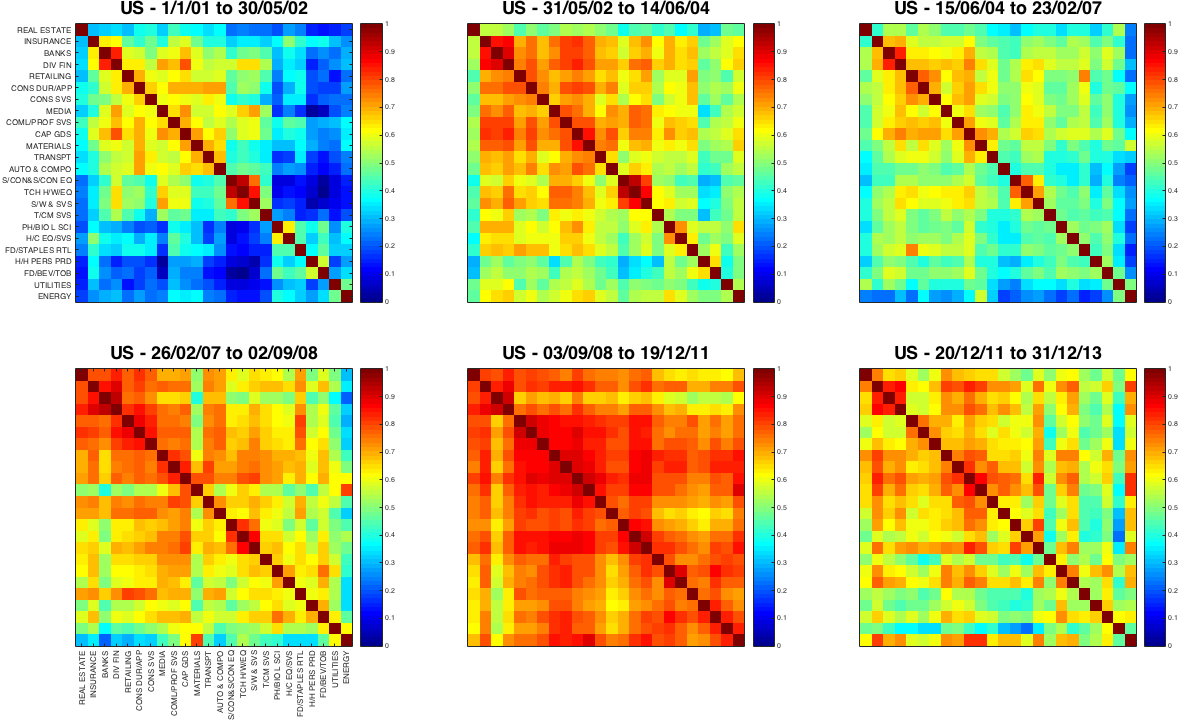}
\caption{Heatmaps of correlation matrix for different scenarios in the US. Ordering is preserved throughout all matrices. Notice this ordering is not the same as for the EMU.} \label{fig::US_Heat}
\end{figure}

\begin{figure}[!htb]
\includegraphics[width=\textwidth]{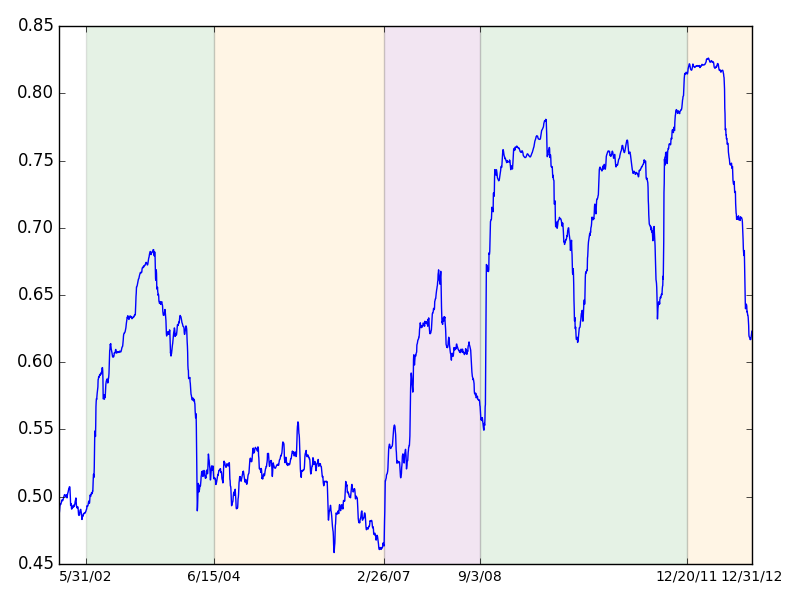}
\caption{Proportion of Variance explained by First Principal Component in the US over the different scenarios.} \label{fig::US_RORO}
\end{figure}

As before, we use the changepoint detection proposed in \cite{changepoint} to come up with our scenarios. We can see the heatmaps of each individual scenario in Figure \ref{fig::US_Heat}. We verify this is sensible by analysing Figure \ref{fig::US_RORO}. The changepoints detected are 31/05/02, 15/06/04, 26/02/07, 03/09/08 and 20/12/11.

We use the 6 covariance matrices for both PRCPO and SARPO. For MSPO we use the covariance matrix calculated over the whole training period as our estimate. For all three portfolio optimisation problems we will use as the expected return of each individual asset the realised return over the most recent scenario -- in this case, the realised return between 20/12/11 and 31/12/12. This is an arbitrary choice that is by no means a superior predictor to any other prediction, but is simple enough as a rule of thumb for us to use it as an example. Not only is forecasting expected returns well beyond the scope of this paper, but they also have a disproportionate impact on portfolios weights when compared to expected volatility or correlation, as demonstrated in \cite{demiguel-garlappi-uppal,michaud}. Hence different forecasting techniques are likely to considerably impact the results that follow. We therefore urge readers to go the extra mile if they are committed to incorporating expected returns in portfolio optimisation problems.

These portfolios' weights can be found in Table \ref{table::US_comp}. It is unequivocal that different choices have been made, despite the fact that the objective function (i.e. the expected returns) is the same for all three portfolios. PRCPO focus more on Food, Beverage $\&$ Tobacco, while SARPO and MSPO focus more on Telecommunication Services, Media and Retailing. They are all fairly diversified, but this is merely a symptom of the imposed cap. PRCPO agrees with both SARPO and MSPO on $60.56\%$, meaning $39.44\%$ of the capital is allocated differently.

\begin{table}[!htpb]
\begin{center}
\begin{tabular}{| l | r | r | r | r |}
\hline
& PRCPO & SARPO & MSPO \\ \hline
Food \& Staples Retailing & 20\% & 20\% & 20\% \\ \hline
Food, Beverage \& Tobacco & 20\% & 0\% & 0\% \\ \hline
Pharmaceuticals, & & & \\
Biotechnology \& Life Sciences & 20\% & 20\% & 20\% \\ \hline
Household \& Personal Products & 19.44\% & 0\% & 0\% \\ \hline
Telecommunication Services & 12.94\% & 20\% & 20\% \\ \hline
Media & 7.62\% & 20\% & 20\% \\ \hline
Retailing & 0\% & 19.12\% & 9.40\% \\ \hline
Real Estate & 0\% & 0\% & 5.93\% \\ \hline
Diversified Financials & 0\% & 0\% & 4.66\% \\ \hline
Health Care Equipment \& Services & 0\% & 0.88\% & 0\% \\ \hline
\end{tabular}
\end{center} \caption{Weights of Maximum Expected Return Portfolio Optimisation with Proportional Regret Constraint (PRCPO), Sharpe Ratio Absolute Robust Portfolio Optimisation (SARPO) and Maximum Sharpe Ratio Portfolio Optimisation (MSPO) for US.} \label{table::US_comp}
\end{table}

\begin{figure}[!htb]
\includegraphics[width=\textwidth]{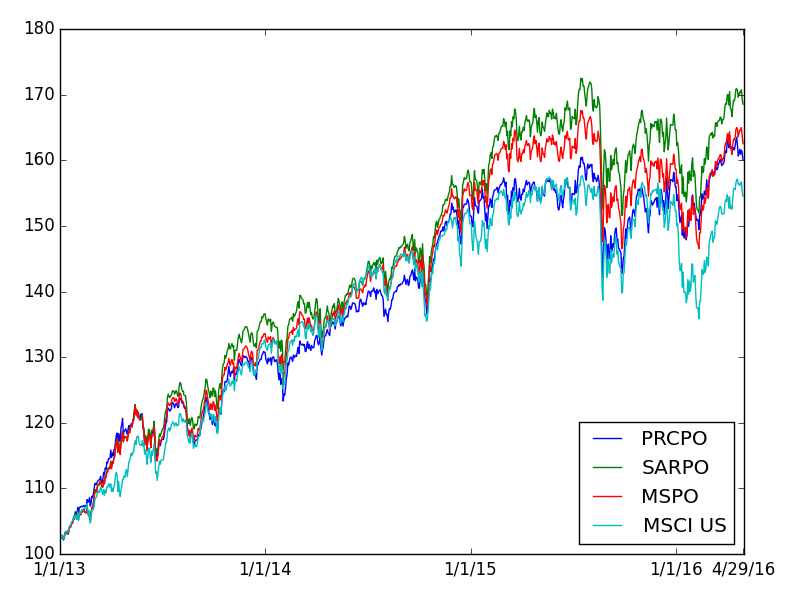}
\caption{Total return of PRCPO, SARPO, MSPO and MSCI US} \label{fig::US_perf}
\end{figure}

We can now analyse how these portfolios perform on the test period, from 2013 until April 2016 both in Figure \ref{fig::US_perf} and in Table \ref{table::US_perf}.

\begin{table}[!htb]
\begin{center}
\begin{tabular}{| l | r | r | r | r |}
\hline
& PRCPO & SARPO & MSPO & MSCI US \\ \hline
Realised Sharpe Ratio 2013 & 2.40 & 2.78 & 2.56 & 2.50 \\ \hline
Realised Sharpe Ratio 2014 & 1.48 & 1.20 & 1.32 & 1.08 \\ \hline
Realised Sharpe Ratio 2015 & 0.20 & 0.33 & 0.17 & 0.08 \\ \hline
Realised Sharpe Ratio 2016 & 0.62 & 0.48 & 0.50 & 0.26 \\ \hline
Realised Sharpe Ratio 2013-16 & 1.18 & 1.21 & 1.14 & 0.97 \\ \hline
Realised Volatility 2013 & 10.80pp & 10.82pp & 10.94pp & 10.88pp \\ \hline
Realised Volatility 2014 & 9.61pp & 11.05pp & 10.69pp & 11.22pp \\ \hline
Realised Volatility 2015 & 13.69pp & 14.52pp & 14.35pp & 15.15pp \\ \hline
Realised Volatility 2016 & 12.29pp & 14.21pp & 14.21pp & 16.45pp \\ \hline
Realised Volatility 2013-16 & 11.59pp & 12.47pp & 12.35pp & 13.02pp \\ \hline
\end{tabular}
\end{center} \caption{Annualised Sharpe ratio and volatility of PRCPO, SARPO, MSPO and MSCI US over different periods.} \label{table::US_perf}
\end{table}

We can see that SARPO has a slightly higher annualised Sharpe ratio, but this however comes at the cost of higher volatility than we would consider admissible (this is relevant as we require the portfolio to be fully invested (\ref{simple}). In $2014$, for example, SARPO and MSPO had respectively $17.2\%$ and $13.4\%$ more volatility than the least volatile benchmark, the Consumer Staples sector group, while PRCPO had only $1.93\%$ extra volatility, well below the $10\%$ cap imposed. While both did outperform the market over this period, it could easily have gone the other way. This is a good example of why we think a proportional regret constraint is important -- in this case it allowed us to pursue a high Sharpe Ratio while safeguarding against taking too much extra volatility.

\subsection*{Results across multiple countries}

As a way of reinforcing the points made earlier we carried out the same experiments on equity data from Australia, Canada, France, Japan, South Korea and United Kingdom. For each we calculated new changepoints and recomputed the portfolios above.

Table \ref{table::summary} in the Appendix summaries the results for each region. In the minimum regret problem we outline the percentage of capital that is allocated differently from RRPO to ARPO and MVPO, the biggest weight each allocates to a single industry group, their realised return and volatility for the entirety of the testing period. In the maximum expected value problem we display the percentage of capital allocated differently from PRCPO to SARPO and MSPO, their Sharpe ratio and volatility (the maximum weight is $20\%$ by construction).

A few important points can be taken from this table. First, it is clear that both RRPO and PRCPO generally offer different solutions not covered by other methods. Secondly, RRPO tends to offer more diversification (by having less weight in a single industry group) than both ARPO and MVPO. In fact, the only instances when it does not is when the amount of capital allocated differently is marginal. This comes naturally from the fact that RRPO takes all scenarios into account, not only the worst-case scenario (as ARPO) and so favours diversification.

Thirdly, PRCPO is the ``worst'' strategy over only one region (Australia), against four for SARPO and three for MSPO. Moreover, the only time one strategy significantly outperforms the others is PRCPO outperforming both SARPO and MSPO in Canada.
Finally, it is worth mentioning that over the past few years low volatility assets have outperformed the market. This means that portfolios that invest in these assets, such as the ones examined, have also outperformed the market. There is no reason to believe this to be indefinitely true, and so none of these portfolios is guaranteed to persistently outperform the market.


\section{Final Remarks}

We demonstrated that regret minimisation does not only produce novel but also sensible results. Indeed, we observed that across different datasets it provides a greater degree of protection than absolute robust optimisation and the more widely used minimum volatility optimisation.
A minimum regret portfolio optimisation problem assumes a fresh start and so should be interesting from an academic point of view, although it may also be implemented by a defensive portfolio manager, whose main concern is not to be too heavily outperformed.
A proportional regret constraint constructed using a suitable set of benchmarks and scenarios, however, would give investors the confidence to pursue their objectives while at the same time controlling the level of extra volatility one is prepared to take. Consequently it is a tool that deserves to be considered by any portfolio manager.

This is clearly not the El Dorado in asset management, but instead a small cog in a much larger (and hopefully successful) engine. The best results will be achieved by those who possess a profitable trading strategy and combine it with a wide range of techniques including but not limited to robust parameter estimation, inclusion of regularisation terms, transaction costs modelling and tail risk modelling.

%% file: appendix.tex
\setlength{\tabcolsep}{4pt}

\begin{table}[!htb]
\begin{center}
\begin{tabular}{| l | c | c | c | c | c | c | c | c |}
\hline
& AUS & CAN & FRA & JPN & KOR & UK & EMU & US \\ \hline
A1 & 30.51\% & 2.33\% & 33.16\% & 33.59\% & 36.32\% & 5.85\% & 29.67\% & 12.62\% \\ \hline
A2 & 20.26\% & 14.36\% & 28.85\% & 14.45\% & 8.97\% & 3.07\% & 23.86\% & 9.54\% \\ \hline \hline

B1 & 25.15\% & 38.96\% & 19.22\% & 25.12\% & 35.49\% & 32.37\% & 35.84\% & 53.22\% \\ \hline
B2 & 29.04\% & 39.74\% & 26.38\% & 43.06\% & 43.60\% & 33.07\% & 50.23\% & 58.37\% \\ \hline
B3 & 28.24\% & 41.83\% & 25.84\% & 30.13\% & 39.26\% & 31.78\% & 44.17\% & 49.31\% \\ \hline \hline

C1 & 9.64\% & 13.30\% & 9.07\% & 19.58\% & 7.54\% & 7.57\% & 14.49\% & 13.82\% \\ \hline
C2 & 12.06\% & 13.23\% & 7.72\% & 20.74\% & 9.61\% & 8.06\% & 14.93\% & 14.27\% \\ \hline
C3 & 10.78\% & 15.10\% & 7.80\% & 19.29\% & 8.27\% & 7.18\% & 14.59\% & 13.96\% \\ \hline 
C4 & 7.65\% & 6.35\% & 9.28\% & 14.23\% & -0.83\% & 4.92\% & 9.11\% & 12.62\% \\ \hline \hline

D1 & 12.88pp & 11.45pp & 17.14pp & 20.02pp & 12.98pp & 12.84pp & 16.33pp & 11.53pp \\ \hline
D2 & 13.16pp & 11.54pp & 16.82pp & 19.73pp & 14.44pp & 12.77pp & 16.95pp & 11.50pp \\ \hline
D3 & 13.03pp & 11.19pp & 16.72pp & 19.93pp & 13.12pp & 12.88pp & 16.73pp & 11.40pp \\ \hline
D4 & 14.28pp & 12.44pp & 18.28pp & 22.49pp & 13.12pp & 14.35pp & 18.05pp & 13.02pp \\ \hline \hline

E1 & 23.21\% & 35.58\% & 24.41\% & 22.24\% & 0.00\% & 15.32\% & 18.49\% & 39.44\% \\ \hline
E2 & 40.00\% & 43.57\% & 11.74\% & 19.00\% & 8.57\% & 15.54\% & 29.51\% & 39.44\% \\ \hline \hline

F1 & 0.99 & 1.04 & 0.59 & 0.92 & 0.38 & 0.85 & 0.74 & 1.18 \\ \hline
F2 & 1.09 & 0.61 & 0.52 & 0.83 & 0.38 & 0.75 & 0.78 & 1.21 \\ \hline
F3 & 1.05 & 0.76 & 0.61 & 0.94 & 0.34 & 0.86 & 0.68 & 1.14 \\ \hline
F4 & 0.54 & 0.51 & 0.51 & 0.63 & -0.06 & 0.34 & 0.50 & 0.97 \\ \hline \hline

G1 & 13.28pp & 11.19pp & 17.20pp & 20.27pp & 14.25pp & 13.19pp & 16.99pp & 11.59pp \\ \hline
G2 & 12.67pp & 16.96pp & 17.67pp & 21.21pp & 14.25pp & 13.63pp & 16.96pp & 12.47pp \\ \hline
G3 & 12.45pp & 16.83pp & 16.83pp & 20.19pp & 13.42pp & 13.89pp & 16.61pp & 12.35pp \\ \hline
G4 & 14.28pp & 12.44pp & 18.28pp & 22.49pp & 13.12pp & 14.35pp & 18.05pp & 13.02pp \\ \hline
\end{tabular}
\end{center}
\caption{Summary of results across 8 different regions, from 2013 until April 2016. \newline
A - Percentage of capital RRPO invests differently from ARPO~(1) and MVPO~(2). \newline
B - Highest weight in RRPO~(1), ARPO~(2) and MVPO~(3). \newline
C - Realised return of RRPO~(1), ARPO~(2), MVPO~(3) and MSCI Index~(4). \newline
D - Realised volatility of RRPO~(1), ARPO~(2), MVPO~(3) and MSCI Index~(4). \newline
E - Percentage of capital PRCPO invests differently from SARPO~(1) and MSPO~(2). \newline
F - Realised Sharpe ratio of PRCPO~(1), SARPO~(2), MSPO~(3) and MSCI Index~(4). \newline
G - Realised volatility of PRCPO~(1), SARPO~(2), MSPO~(3) and MSCI Index~(4).
} \label{table::summary}
\end{table}